\documentclass{article}
\usepackage{amssymb}
\usepackage{amsfonts}
\usepackage{amsmath}

\newtheorem{theorem}{Theorem}

\newtheorem{definition}[theorem]{Definition}

\newtheorem{proposition}[theorem]{Proposition}
\newtheorem{remark}[theorem]{Remark}

\input{tcilatex}

\begin{document}

\title{New considerations on Einstein equations in anisotropic spaces}
\author{Nicoleta VOICU \\
"Transilvania" University, Brasov, Romania}
\maketitle

\begin{abstract}
We find the generalization of Einstein equations to Finsler spaces by
variational means and, based on the invariance of the Finslerian Hilbert
action to infinitesimal transformations, we find the analogous of the
energy-momentum conservation law in these spaces.
\end{abstract}

\section{Introduction}

As already shown by meany researchers, Finsler geometry is an important
alternative to be taken into account for a theory of gravitational field
which could solve some important problems of modern astrophysics, such as,
for instance, \cite{agd}: the rotation curves of spiral galaxies, the
3D-problem for spiral galaxies (usual gravity theory doesn't work in the
plane of the galaxy but works in the orthogonal direction), or the location
of globular clusters (which is close to the center of the galaxy and not on
its periphery, as expected) etc.

In this context, finding a generalization of Einstein equations to Finsler
spaces is a necessary and natural step. Several different variants have
already been proposed.

Thus, R. Miron and M. Anastasiei\textbf{\ }(\cite{Lagrange}, 1984) proposed
a generalization of Einstein equations for $hv$-metrics on the tangent
bundle $TM:$%
\begin{equation*}
G=g_{ij}(x,y)dx^{i}\otimes dx^{j}+v_{ab}(x,y)\delta y^{a}\otimes \delta
y^{b}.
\end{equation*}

On the manifold $TM$ such a metric is Riemannian, which suggests by analogy
the following form for the required generalization:%
\begin{eqnarray*}
&&R_{ij}-\dfrac{1}{2}(R+S)g_{ij}=\mathcal{XT}_{ij},~\ \overset{1}{P}\overset{%
}{_{bj}}=\mathcal{XT}_{bj}, \\
&&\overset{2}{P}\overset{}{_{jb}}=-\mathcal{XT}_{jb},~\ \ S_{bc}-\dfrac{1}{2}%
(R+S)v_{bc}=\mathcal{XT}_{ab}.
\end{eqnarray*}%
where $\mathcal{R}_{\alpha \beta }=(R_{ij},\overset{1}{P}\overset{}{_{ia}},%
\overset{2}{P}\overset{}{_{ai}},S_{ab})$ are the local components of the
curvature tensor of the canonical metrical linear connection on $TM$ (whose
torsion tensor does not vanish) and $\mathbb{T}_{\alpha \beta }$ are the
components of the generalized energy-momentum tensor. The number of
equations in this theory is $4\dim M.$

\bigskip

Later, S. Rutz proposed in\textbf{\ }1993, starting from the geodesic
deviation equations in (pseudo-)Finsler spaces, 
\begin{equation*}
\dfrac{Dw^{i}}{ds^{2}}=(R_{j~kl}^{~i}y^{j}y^{l})w^{k},
\end{equation*}%
as generalization of Einstein equations in vacuum in these spaces: 
\begin{equation*}
H_{k}^{~i}:=R_{j~kl}^{~i}y^{j}y^{l}=0.
\end{equation*}

This is an intuitive approach, pointing out the importance of the \textit{%
deviation operator} $H_{k}^{i},$ \cite{Rutz}.

More recently, in a series of three papers, (2007, 2008, 2009), Xin Li, Zhe
Chang, proposed another generalization for a special class of Finsler
spaces, namely, for Berwald spaces. In the cited papers, they started from
Akbar-Zadeh's definition of Ricci tensor:%
\begin{equation*}
Ric_{ik}=\dfrac{1}{2}\dfrac{\partial ^{2}}{\partial y^{i}\partial y^{k}}%
(R_{j~hl}^{~h}y^{j}y^{l}).
\end{equation*}%
and from Bianchi identities for the Chern connection), thus getting:

\begin{equation*}
\{Ric_{jl}-\dfrac{1}{2}g_{jl}S\}+\{\dfrac{1}{2}B_{h~jl}^{~h}+B_{j~lh}^{~h}%
\}=8\pi GT_{jl},
\end{equation*}

where 
\begin{equation*}
S=g^{ij}R_{ij},~\ \ B_{ijkl}=-\dfrac{1}{2}g_{ij\cdot u}R_{~kl}^{u}.
\end{equation*}

All the above approaches provide different variants, with different ranges
of applicability. Moreover, none of them is based on variational methods.

A first variational approach for a gravitational field equation in
Finslerian spaces is proposed by G. I. Garas'ko (2009), by taking a
Lagrangian which is proportional to the inverse of the volume of the
indicatrix of a pseudo-Finslerian space. Still, the obtained field equations%
\textbf{\ }do not provide Einstein equations in the particular case of
pseudo-Riemannian spaces.

\bigskip

In this paper, we propose a new generalization of Einstein equations to
Finsler space, based entirely on variational approaches and also a
generalization of the energy conservation law, starting from the invariance
of our generalized Hilbert type action to infinitesimal diffeomorfisms.

\section{Pseudo-Finslerian spaces}

Let $M$ be a 4-dimensional differentiable manifold of class $\mathcal{C}%
^{\infty },$ $(TM,\pi ,M)$ its tangent bundle and $(x^{i},y^{i})_{i=%
\overline{1,4}}$ the coordinates in a local chart on $TM.$ By "smooth" we
shall always mean $\mathcal{C}^{\infty }$-differentiable.

A \textit{pseudo-Finslerian function} on $M,$ is a function $\mathcal{F}%
:TM\rightarrow \mathbb{R}$ with the properties:

\begin{enumerate}
\item $\mathcal{F=F}(x,y)$ is smooth for $y\not=0;$

\item $\mathcal{F}$ is positive homogeneous of degree 1, i.e., $\mathcal{F}%
(x,\lambda y)=\lambda \mathcal{F}(x,y)$ for all $\lambda >0;$

\item The \textit{pseudo-Finslerian metric tensor}:\textit{\ } 
\begin{equation}
g_{ij}(x,y)=\dfrac{1}{2}\dfrac{\partial ^{2}\mathcal{F}^{2}}{\partial
y^{i}\partial y^{j}}  \label{Finsler_metric}
\end{equation}%
is nondegenerate: $\det (g_{ij}(x,y))\not=0,~\forall x\in M,$ $y\in
T_{x}M\backslash \{0\}.$
\end{enumerate}

\bigskip

In the literature, pseudo-Finslerian spaces are also sometimes called simply 
\textit{Finslerian.}

The equations of geodesics $s\mapsto (x^{i}(s))$ of a Finsler space $(M,%
\mathcal{F})$, 
\begin{equation*}
\dfrac{dy^{i}}{ds}+2G^{i}(x,y)=0,~\ y^{i}=\dot{x}^{i},
\end{equation*}%
give rise to an Ehresmann (nonlinear) connection on $TM,$ called the \textit{%
Cartan nonlinear connection}, of local coefficients $\overset{c}{N}\overset{}%
{_{~j}^{i}}=\dfrac{\partial G^{i}}{\partial y^{j}}.$ Let%
\begin{equation*}
\delta _{i}=\dfrac{\partial }{\partial x^{i}}-\overset{c}{N}\overset{}{%
_{~i}^{a}}\dfrac{\partial }{\partial y^{a}},~\ \ \dot{\partial}_{a}=\dfrac{%
\partial }{\partial y^{a}}
\end{equation*}%
be the corresponding adapted basis and 
\begin{equation*}
(dx^{i},~\delta y^{a}=dy^{a}+G_{~i}^{a}dx^{i}),
\end{equation*}%
its dual basis. Then, the equations of geodesics of $(M,\mathcal{F})$ become 
$\dfrac{\delta y^{i}}{ds}=0,~y^{i}=\dot{x}^{i},~\ i=1,...,4.$

We can also obtain other nonlinear connections $N$ on $TM$ by adding to $(%
\overset{c}{N}\overset{}{_{~i}^{a}})$ the components of a (1,1)-type tensor
field :%
\begin{equation}
N_{~i}^{a}=\overset{c}{N}\overset{}{_{~i}^{a}}+X_{~i}^{a}.  \label{nonlinear}
\end{equation}

\bigskip

Any vector field $V$ on $TM$ can be written as%
\begin{equation*}
V=V^{i}\delta _{i}+\tilde{V}^{a}\dot{\partial}_{a};
\end{equation*}%
the component $hV=V^{i}\delta _{i}$ is a vector field, called the \textit{%
horizontal }component of $V,$ while $vV=\tilde{V}^{a}\dot{\partial}_{a}$ is
its \textit{vertical }component. Similarly, a 1-form $\omega $ on $TM$ can
be decomposed as $\omega =\omega _{i}dx^{i}+\tilde{\omega}_{a}\delta y^{a},$
with $h\omega =\omega _{i}dx^{i}$ called the \textit{horizontal} component,
and $v\omega =\tilde{\omega}_{a}\delta y^{a}$ the \textit{vertical }one%
\textit{.}

We shall usually denote indices corresponding to horizontal geometrical
objects by \thinspace $i,j,k,...$, indices corresponding to vertical ones by 
$a,b,c...$ and $\alpha ,\beta ,\gamma ,...\in \{i,j,k,...,a,b,c...\}$ will
denote those which can correspond to either of the distributions.

\bigskip

The \textit{Chern-type linear connection }$C\Gamma (N)=(F_{~jk}^{i},0)$ 
\textit{on }$TM$ has the local coefficients: 
\begin{equation}
F_{~jk}^{i}=\dfrac{1}{2}g^{ih}(\delta _{k}g_{hj}+\delta _{j}g_{hk}-\delta
_{h}g_{jk}).  \label{Chern}
\end{equation}%
We denote by $_{|i}$ and $_{\cdot i}$ the corresponding covariant
derivations 
\begin{eqnarray*}
W_{~|i}^{\alpha } &=&\delta _{i}W^{\alpha }+F_{~ki}^{j}W^{\alpha },~ \\
W_{~\cdot a}^{\alpha } &=&\dfrac{\partial W^{\alpha }}{\partial y^{a}},
\end{eqnarray*}%
(where $W^{\alpha }$ are local coordinates of a vector field $W$ on $TM$).

There holds%
\begin{equation*}
g_{ij|k}=0,~\ \ \forall i,j,k=1...4.
\end{equation*}

\bigskip

The curvature tensor $R(X,Y)Z=D_{X}D_{Y}Z-D_{Y}D_{X}Z-D_{[X,Y]}Z$ has the
following essential local components:

\begin{equation*}
R(\delta _{l},\delta _{k})\delta _{j}=R_{j~kl}^{~i}\delta _{i},~\ R(\dot{%
\partial}_{l},\delta _{k})\delta _{j}=P_{j~kl}^{~i}\delta _{i},
\end{equation*}%
where

\begin{eqnarray*}
R_{~j~kl}^{i} &=&\delta _{l}F_{~jk}^{i}-\delta
_{k}F_{~jl}^{i}+F_{~jk}^{h}F_{~hl}^{i}-F_{~jl}^{h}F_{~hk}^{i}, \\
P_{j~kl}^{~i} &=&F_{~jk\cdot l}^{i}.
\end{eqnarray*}

\bigskip

Geodesics $s\mapsto x^{i}(s)$ on space-time $M$ are characterized by%
\begin{equation}
\dfrac{Dy^{i}}{ds}\equiv \dfrac{dy^{i}}{ds}+F_{~jk}^{i}y^{j}y^{k}=0,
\label{geodesics}
\end{equation}%
and their deviations, by%
\begin{equation}
\dfrac{D^{2}w^{i}}{ds}=R_{j~kl}^{~i}y^{j}y^{l}.  \label{deviations}
\end{equation}

\bigskip

\begin{remark}
The above relations (\ref{geodesics}) and (\ref{deviations}) do not depend
on the nonlinear connection we use. Namely, if we choose some other
nonlinear connection of coefficients $N_{~j}^{i}$ and $F_{~jk}^{i}$ are of
Chern type (\ref{Chern}), then geodesics and their deviations are still
described by these relations. Hence, it appears as convenient to consider
momentarily an arbitrary nonlinear connection $N=(N_{~j}^{i}).$
\end{remark}

Also, in our further considerations, the following quantities:%
\begin{equation*}
C_{ijk}=\dfrac{1}{2}g_{ij\cdot k}
\end{equation*}%
(components of the \textit{Cartan tensor}) will be important.

\bigskip

\section{h-v metric structures on $TM.$ Divergence of vector fields on $TM$}

A tensor on $TM$ is called \textit{distinguished }(or \textit{Finslerian}), 
\cite{Lagrange}, if, with respect to coordinate changes on $TM$ induced by
coordinate changes $(x^{i})\rightarrow (x^{i^{\prime }})$ on $M,$ its local
components transform by the same rule as those of a tensor on $M:$

\begin{equation*}
\mathcal{T}_{~\ \ \ \ \ \ \ \ j_{1}^{\prime }....j_{p}^{\prime
}}^{i_{1}^{\prime }....i_{k}^{\prime }}(x,y)=\dfrac{\partial
x^{i_{1}^{\prime }}}{\partial x^{i_{1}}}...\dfrac{\partial x^{i_{k}^{\prime
}}}{\partial x^{i_{k}}}\dfrac{\partial x^{j_{1}}}{\partial x^{j_{1}^{\prime
}}}...\dfrac{\partial x^{j_{k}}}{\partial x^{j_{k}^{\prime }}}\mathcal{T}%
_{~~\ ~\ ~\ \ \ j_{1}...j_{p}}^{i_{1}....i_{k}}.
\end{equation*}

For instance, Finslerian metric tensors (\ref{Finsler_metric}) are
distinguished tensors on $TM.$ Also, the elements $\delta _{i},\dot{\partial}%
_{a}$ are distinguished vector fields, while $dx^{i},\delta y^{a}$ are
distinguished covector fields.

\begin{definition}
An h-v metric on $TM$ is a metric structure of the form 
\begin{equation}
G_{\alpha \beta }(x,y)=g_{ij}(x,y)dx^{i}\otimes dx^{j}+v_{ab}(x,y)\delta
y^{a}\otimes \delta y^{b}  \label{h-v metric}
\end{equation}%
where $g_{ij}$ and $v_{ab}$ denote (0,2)-type distinguished tensors with $%
\det (g_{ij})\not=0,$ $\det (v_{ab})\not=0.$
\end{definition}

\begin{remark}
An h-v metric structure is actually a pseudo-Riemannian metric on the
manifold $TM.$
\end{remark}

Let us denote%
\begin{eqnarray*}
g &=&\det (g_{ij}),v=\det (v_{ab}), \\
G &=&\det (G_{\alpha \beta })=gv.
\end{eqnarray*}

The (Riemannian) volume element on $TM$ is%
\begin{equation*}
dV=\ast (1)
\end{equation*}%
(where $\ast $ denotes the Hodge star operator for differential forms). That
is,%
\begin{equation}
dV=\sqrt{|G|}dx^{i}\wedge dx^{2}\wedge ...\wedge \delta y^{4}.
\label{volume_el}
\end{equation}

For simplicity, we shall denote in the following, $dx^{i}\wedge dx^{2}\wedge
...\wedge \delta y^{4}=:d\Omega ,$ hence%
\begin{equation*}
dV=\sqrt{|G|}d\Omega .
\end{equation*}

\bigskip

The \textit{divergence} of a vector field $X=X^{i}\delta _{i}+\tilde{X}^{a}%
\dot{\partial}_{a}\in \mathcal{X}(TM)$ is defined as $div(X)=d(\ast X^{b}),$
where $^{b}$ denotes the musical isomorphism (lowering indices). In local
coordinates,%
\begin{equation*}
div(X)=div(X^{H})+div(X^{V}),
\end{equation*}%
where%
\begin{eqnarray}
div(X^{H}) &=&X_{~~|i}^{i}-P_{i}X^{i},~\ \ P_{i}=N_{~i\cdot a}^{a}-\delta
_{i}(\ln \sqrt{\mathcal{G}}),  \label{divergence_h} \\
divX^{V} &=&\tilde{X}_{~\cdot a}^{a}+\tilde{P}_{a}\tilde{X}^{a},~\ \tilde{P}%
_{a}=\dot{\partial}_{a}(\sqrt{G})  \label{divergence_v}
\end{eqnarray}%
(and $_{|i}$ denotes covariant derivative with respect to the Chern-type
linear connection).

\bigskip

\section{Hilbert action and Einstein equations}

Our aim in the following is to define a Hilbert action for Finslerian
spaces, which should:

- be as simple as possible and yield as simple equations as possible;

- in the particular case of Riemannian spaces, provide the regular Einstein
equations;

\bigskip

Hilbert proposed as a "simplest scalar" (in the pseudo-Riemannian case):%
\begin{equation*}
R=g^{ij}R_{ij}.
\end{equation*}

This "simplest scalar" \textit{characterizes geodesic deviation}, in the
following sense:%
\begin{equation}
geodesic~~deviation:~~\dfrac{Dw^{i}}{ds^{2}}=H_{~k}^{i}w^{k},~\ ~\ \ \ \ \
H_{~k}^{i}=R_{j~kl}^{~i}y^{j}y^{l},~\ \ y^{i}=\dfrac{dx^{i}}{ds}.
\label{geodesic_deviation}
\end{equation}

Then%
\begin{equation*}
R_{jl}y^{j}y^{l}=trace(H_{~l}^{i})
\end{equation*}

In particular, Einstein equations in vacuum can be written as:%
\begin{equation}
H_{k}^{~i}=0.  \label{R_Einstein}
\end{equation}%
(which justifies Rutz's intuitive approach).

\bigskip

\textbf{In the Finslerian case, }we can think of the following possibilities:

1. by means of Ricci tensor $Ric_{ij}=\dfrac{1}{2}\dfrac{\partial
^{2}H_{l}^{l}}{\partial y^{i}\partial y^{j}}$ \ as defined by Akbar-Zadeh;

2. by means of Chern, Cartan or Berwald connection curvature?;\ 

3. eventually add the contracted vertical curvature $S=g^{ij}S_{ij}$ (for
Cartan connection) and get $R+S$ as "simplest scalar" (as suggested by Miron
and Anastasiei's approach).

\bigskip

We notice that, among the above variants,\textbf{\ }$R_{jk}=R_{j~ki}^{~i}$
built from the curvature of the Chern connection provides the simplest
computations, while the intuitive interpretation (\ref{geodesic_deviation})
and (\ref{R_Einstein}) is satisfied.

\bigskip

There is still a technical problem, represented by the non-holonomy of the
frame $(\delta _{i},\dot{\partial}_{a}),$ which yields extra terms when
performing variation - expectedly leading to complicated equations. In order
to solve this problem, we propose the use of another nonlinear connection
than Cartan's. Thus, the intuitive interpretation (\ref{geodesic_deviation})
and (\ref{R_Einstein}) will be still sufficed, and .the obtained equations
will acquire a simple form.

Briefly, we shall perform the following steps:

\begin{enumerate}
\item Define an analogue of Hilbert action, for anisotropic spaces;\textbf{\ 
}

\item Find a most convenient nonlinear connection (appropriate frame on $TM$%
).

\item Variate the Hilbert action w.r.t. the metric tensor\textit{\ }$g_{ij}\ 
$and get the Einstein equations;

\item Find the analogue of the conservation law for the energy-momentum
tensor (some identity verified by the divergence of energy--momentum tensor).
\end{enumerate}

\bigskip

\subsection{Step 1: define Hilbert action}

Let \ 
\begin{equation}
G_{\alpha \beta }(x,y)=g_{ij}(x,y)dx^{i}\otimes dx^{j}+v_{ab}(x,y)\delta
y^{a}\otimes \delta y^{b},
\end{equation}%
be an h-v metric on $TM,$ such that:

- $g_{ij}$ defines a pseudo-Finslerian metric on $M,$ of signature $%
(+,-,-,-).$

-~$v_{ab}$ denotes an arbitrary and fixed (0,2) distinguished tensor on $TM$%
, of constant signature. For convenience, let us suppose $v=\det (v_{ab})<0,$
which entails $G>0.$

Let

\begin{equation}
R_{jk}=R_{j~ik}^{~i},~\ \ R=R^{jk}g_{jk}  \label{Chern-Ricci}
\end{equation}%
where $%
R_{~jkl}^{~i}=F_{~jk;l}^{i}-F_{~jl;k}^{i}+F_{~jk}^{h}F_{~hl}^{i}-F_{~jl}^{h}F_{~hk}^{i} 
$ is the curvature of Chern connection.

\textbf{Interpretation:} $R_{jk}y^{j}y^{k}$ is the trace (with a minus sign)
of the deviation operator (cf. Rutz):$\ \
H_{~l}^{i}=R_{j~lk}^{~i}y^{j}y^{k}. $

\bigskip

We define the Finslerian Hilbert action as%
\begin{equation}
S=\int R\sqrt{G}d\Omega ,  \label{Hilbert}
\end{equation}%
where integration is made upon some fixed domain $D\subset TM$ in the
tangent bundle. We have: $S=S(g_{ij},~N_{~j}^{i},~g_{ij;k},~g_{ij;k;l}).$

\subsection{Step 2: find a convenient nonlinear connection}

It is convenient for further purposes to choose, if possible, a nonlinear
connection for which $P_{i}=0,$ that is, $N_{~i\cdot a}^{a}-\delta _{i}(\ln 
\sqrt{-v})=0.$ According to (\ref{divergence_h}), in this case, the
divergence of a horizontal vector field $X^{H}$ on $TM$ will be written
simply as:%
\begin{equation}
div(X^{H})=X_{~~|i}^{i}.  \label{div_simplified}
\end{equation}

\bigskip

Hence, let us consider:%
\begin{equation*}
N_{~i}^{a}=G_{~i}^{a}+y^{a}V_{i},
\end{equation*}%
where $V_{j}=V_{j}(x,y)$ are the components of a 0-homogeneous in $y$
distinguished vector field to be determined.

By imposing the condition $N_{~i\cdot a}^{a}-\delta _{i}(\ln \sqrt{\mathcal{G%
}})=0,$ we get%
\begin{equation*}
\delta _{i}(\ln \sqrt{\mathcal{G}})=~\overset{c}{N}\overset{}{_{~i\cdot
a}^{a}}+y_{~\cdot a}^{a}V_{i}+y^{a}V_{i\cdot a}.
\end{equation*}%
By 0-homogeneity of $V_{i}$, it follows that $y^{a}V_{i\cdot a}=0,$ which
leads to%
\begin{equation*}
V_{i}=\dfrac{1}{4}(\delta _{i}(\ln \sqrt{\mathcal{G}})-~\overset{c}{N}%
\overset{}{_{~i\cdot a}^{a}})
\end{equation*}%
and%
\begin{equation}
N_{~i}^{a}=G_{~i}^{a}+\dfrac{1}{4}(\delta _{i}(\ln \sqrt{\mathcal{G}})-~%
\overset{c}{N}\overset{}{_{~i\cdot a}^{a}})  \label{N}
\end{equation}

By direct computation, it can be proven that $N_{~i}^{a}$ obey the rule of
transformation of the coefficients of a nonlinear connection (\cite{Lagrange}%
), with respect to coordinate changes on $TM$.

Moreover, for this nonlinear connetion, the Chern-type connection
coefficients are%
\begin{equation*}
L_{~jk}^{i}=\dfrac{1}{2}g^{ih}(g_{hj~\tilde{;}~k}+g_{hk~\tilde{;}~j}-g_{jk~%
\tilde{;}~h})=F_{~jk}^{i}-\dfrac{1}{2}g^{ih}(y^{l}V_{k}g_{hj\cdot
l}+y^{l}V_{j}g_{hk\cdot l}-y^{l}V_{h}g_{jk\cdot l}).
\end{equation*}%
By the 0-homogeneity of $g_{ij},$ the terms in the last bracket vanish,
which means that $L_{~jk}^{i}=F_{~jk}^{i}.$ Thus, there holds

\begin{proposition}
\begin{enumerate}
\item The functions (\ref{N}) are the local coefficients of a nonlinear
connection on $TM.$

\item The coefficients $L_{~jk}^{i}=\dfrac{1}{2}g^{ih}(g_{hj~\tilde{;}%
~k}+g_{hk~\tilde{;}~j}-g_{jk~\tilde{;}~h})$ of the corresponding Chern type
connection coincide with the usual Chern connection coefficients $%
F_{~jk}^{i}.$
\end{enumerate}
\end{proposition}

\bigskip

That is, we can formally use nonlinear connection (\ref{N}) instead of the
usual Cartan one, without changing either the Ricci scalar or the Ricci
tensor, but having (\ref{div_simplified}).

\subsection{Step 3: variation w.r.t. the metric $g_{ij}$, Einstein equations}

Now, having the nonlinear connection (\ref{N}) and Chern covariant
derivation (\ref{Chern}), let us perform the variation with respect to the
metric\textbf{\ }$g_{ij}$ of the Finslerian Hilbert action (for vacuum
case). That is, we shall variate the horizontal part of the hv-metric
structure $G_{\alpha \beta }$ and keep its vertical part $v_{ab}$ fixed.

We get%
\begin{eqnarray*}
\mathbf{\delta }_{g}S &=&\int \mathbf{\delta }(g^{ij}R_{ij}\sqrt{G})d\Omega =
\\
&=&\int R_{ij}\mathbf{\delta }g^{ij}\sqrt{G}d\Omega ~+~\int g^{ij}\mathbf{%
\delta }(R_{ij})\sqrt{G}d\Omega ~+\int g^{ij}R_{ij}\mathbf{\delta (}\sqrt{G}%
)d\Omega .
\end{eqnarray*}

By means of relation (\ref{div_simplified}), we get that the divergence of a
horizontal vector field $V=V^{i}\delta _{i}$ on $TM$ can be written simply
as $div(V)=V_{~|i}^{i}\sqrt{-g}$ , and the second integral is%
\begin{equation*}
\int g^{ij}\mathbf{\delta }(R_{ij})\sqrt{-g}d\Omega =\int \{g^{jk}(\mathbf{%
\delta }F_{~ij}^{i})-g^{ij}(\mathbf{\delta }F_{~ij}^{k})\}_{|k}\sqrt{-g}%
d\Omega =\int (V^{k}\sqrt{-g})_{,k}d\Omega =0
\end{equation*}%
(we suppose that we can make the variations $\mathbf{\delta }F_{~jk}^{i}$
vanish on the boundary of the domain of integration).

The third integral is 
\begin{equation*}
\int g^{ij}R_{ij}\mathbf{\delta (}\sqrt{G})d\Omega =\int g^{ij}R_{ij}\mathbf{%
\delta (}\sqrt{-g})\sqrt{-\mathcal{G}}d\Omega =-\dfrac{1}{2}\int Rg_{ij}%
\sqrt{G}d\Omega .
\end{equation*}

\bigskip

We get:%
\begin{equation*}
\mathbf{\delta }_{g}S=\int (R_{ij}-\dfrac{1}{2}Rg_{ij})\mathbf{\delta }g^{ij}%
\sqrt{-g}d\Omega ,
\end{equation*}%
which yields the \textit{Einstein equations in vacuum for the Finsler space }%
$(M,F):$%
\begin{equation}
R_{ij}-\dfrac{1}{2}Rg_{ij}=0.  \label{vacuum_eqns}
\end{equation}

By adding to the action some term $\mathcal{L}_{matter},$ the corresponding
stress-energy tensor is%
\begin{equation*}
\dfrac{\mathbf{\delta }\mathcal{L}_{matter}}{\mathbf{\delta }g_{ij}}=%
\mathcal{X}T_{ij},
\end{equation*}%
(where $\mathcal{X}$ is a constant), and we get

\begin{theorem}
The Einstein field equations in the Finsler space $(M,F)$ are%
\begin{equation}
R_{ij}-\dfrac{1}{2}Rg_{ij}=\mathcal{X}T_{ij}.  \label{Einstein}
\end{equation}
\end{theorem}

\bigskip

The quantity%
\begin{equation*}
\mathcal{G}_{ij}\equiv R_{ij}-\dfrac{1}{2}Rg_{ij}=\dfrac{\mathbf{\delta }%
\mathcal{L}_{Hilbert}}{\mathbf{\delta }g_{ij}}
\end{equation*}%
is called the \textit{Einstein tensor} for the Finsler space $(M,F).$ With
this notation, Einstein equations read as%
\begin{equation*}
\mathcal{G}_{ij}=\mathcal{X}T_{ij}.
\end{equation*}

\bigskip

\subsection{Step 4: diffeomorphism invariance of Finslerian Hilbert action
and energy-momentum conservation}

The correct way of posing the problem of energy momentum conservation, is
deducing it from the invariance to (infinitesimal) diffeomorphisms of
Hilbert action, \cite{Bertschinger}.

An infinitesimal transformation on $M$ induces an \textit{infinitesimal
transformation on} $TM$ as%
\begin{eqnarray*}
\tilde{x}^{i} &=&x^{i}+\varepsilon \xi ^{i}(x), \\
\tilde{y}^{i} &=&y^{i}+\varepsilon \xi _{~,j}^{i}(x)y^{j},
\end{eqnarray*}%
where $\xi =\xi (x)$ is an arbitrary vector field on the base manifold $M.$

We can regard variations $\mathbf{\delta }g_{ij},$ generated by
infinitesimal transformations. Then, the variation of the metric tensor is
given by%
\begin{equation*}
\mathbf{\delta }g_{ij}\equiv \mathcal{L}_{\xi }g_{ij}=\xi _{i|j}+\xi
_{j|i}+2\xi _{~|l}^{k}y^{l}C_{ijk}.
\end{equation*}

\bigskip

It can be easily checked (similarly to \cite{Bertschinger}), that:

\begin{remark}
A\textit{ny scalar action on }$TM$\textit{\ is invariant to infinitesimal
diffeomorphisms}.
\end{remark}

\bigskip

That is, with respect to such transformations, we shall have for the
Finslerian Hilbert action%
\begin{equation*}
\mathbf{\delta }S=\int \mathcal{G}_{ij}\mathbf{\delta }g^{ij}\sqrt{G}d\Omega
=0.
\end{equation*}

(the above is independent on the fact that the metric extremizes the action
or not!) It is more convenient to write it as%
\begin{equation*}
\int \mathcal{G}^{ij}\mathbf{\delta }g_{ij}\sqrt{G}d\Omega =0
\end{equation*}%
Replacing $\delta g_{ij}$ and integrating by parts, we get%
\begin{equation*}
0=-\int \xi ^{k}\{\mathcal{G}_{k|j}^{~j}+(y^{l}\mathcal{G}%
^{ij}C_{ijk})_{|l}\}\sqrt{G}d\Omega .
\end{equation*}

Since the vector field $\xi $ is arbitrary and $\mathcal{G}_{ij}=\mathcal{X}%
T_{ij}$ for solutions of Einstein equations we have proved

\bigskip

\begin{theorem}
The energy-momentum tensor in Finsler spaces satisfies the identities%
\begin{equation}
T_{~~k|j}^{j}+(y^{l}T^{ij}C_{ijk})_{|l}=0,~\ \ k=1,...,4.  \label{sem_tensor}
\end{equation}
\end{theorem}

The above is the way that the "energy conservation" relation $%
T_{~~k|j}^{j}=0 $ (which is not even in the Riemannian case a true
conservation law, since $T_{~~k|j}^{j}=0$ does not entail $divT=0$)
translates to Finsler spaces.

\bigskip

\section{Examples}

\textbf{1)}\ For \textbf{weak metrics }%
\begin{equation*}
g_{ij}(x,y)=\eta _{ij}+\varepsilon _{ij}(x,y),
\end{equation*}%
where $\eta _{ij}=diag(1,-1,-1,-1)$ is the Minkowski metric and $\varepsilon
_{ij}$ is a small deformation ($(\varepsilon _{ij})^{2}\simeq 0$),
Chern-type connection coefficients reduce to regular Christoffel symbols
(with respect to $x$)%
\begin{equation*}
F_{~jk}^{i}\simeq \Gamma _{~jk}^{i},
\end{equation*}%
hence \textit{Einstein equations formally look as in the Riemannian case}.
In the Lorentz gauge (\cite{Carroll}), they reduce to%
\begin{equation*}
\square \varepsilon _{ij}\equiv \eta ^{kl}\varepsilon _{ij,kl}=0.
\end{equation*}

Relations (\ref{sem_tensor}) reduce to%
\begin{equation*}
T_{~~k,j}^{j}=0,
\end{equation*}%
(regular divergence of the energy-momentum tensor vanishes), hence
energy-momentum tensor is conserved.

\textbf{2)} For \textbf{weak conformal deformations }%
\begin{equation*}
g_{ijkl}=\varepsilon (x)\gamma _{ijkl},
\end{equation*}%
\textbf{of the Berwald-Moor metric}, the Einstein equations become again%
\begin{equation*}
\square \varepsilon =0.
\end{equation*}

\textbf{Acknowledgment: }The work was supported by the grant No. 4 /
03.06.2009, between the Romanian Academy and Politehnica University of
Bucharest.

\end{document}